\documentclass{article}

\usepackage{PRIMEarxiv}

\usepackage[utf8]{inputenc} 
\usepackage[T1]{fontenc}    
\usepackage{hyperref}       
\usepackage{url}            
\usepackage{booktabs}       
\usepackage{amsfonts}       
\usepackage{nicefrac}       
\usepackage{microtype}      
\usepackage{lipsum}
\usepackage{fancyhdr}       
\usepackage{graphicx}       
\graphicspath{{media/}}
\usepackage{listings}
\usepackage{makecell}
\usepackage{multirow}

\usepackage{subfigure}
\usepackage{pifont}
\newcommand{\cmark}{\ding{51}}  
\newcommand{\xmark}{\ding{55}}  

\title{Optimizing Code Embeddings and ML Classifiers for Python Source Code Vulnerability Detection
 
}

\author{
  Talaya Farasat \\
  University of Passau \\
  tf@sec.uni-passau.de \\
   \And
  Joachim Posegga \\
   University of Passau \\
  jp@sec.uni-passau.de \\
}

\begin{document}
\maketitle

\begin{abstract}
 In recent years, the growing complexity and scale of source code have rendered manual software vulnerability detection increasingly impractical. To address this challenge, automated approaches leveraging machine learning and code embeddings have gained substantial attention.
 This study investigates the optimal combination of code embedding techniques and machine learning classifiers for vulnerability detection in Python source code. We evaluate three embedding techniques, i.e., Word2Vec, CodeBERT, and GraphCodeBERT alongside two deep learning classifiers, i.e., Bidirectional Long Short-Term Memory (BiLSTM) networks and Convolutional Neural Networks (CNN). While CNN paired with GraphCodeBERT exhibits strong performance, the BiLSTM model using Word2Vec consistently achieves superior overall results. These findings suggest that, despite the advanced architectures of recent models like CodeBERT and GraphCodeBERT, classical embeddings such as Word2Vec, when used with sequence-based models like BiLSTM, can offer a slight yet consistent performance advantage. The study underscores the critical importance of selecting appropriate combinations of embeddings and classifiers to enhance the effectiveness of automated vulnerability detection systems, particularly for Python source code.
\end{abstract}


\section{Introduction}
Software vulnerabilities are a fundamental reason for the prevalence of cyberattacks, and their identification remains a crucial yet challenging problem in cybersecurity. Attackers can compromise the integrity of applications or systems by exploiting weaknesses embedded in the source code.

Traditional software vulnerability detection tools rely on static \cite{b1} and dynamic \cite{b2} code analysis. Static code analysis is typically performed using rule-based systems that depend on predefined features of vulnerable code. Unfortunately, these systems tend to generate a significant number of false positives \cite{b3, b4}. Additionally, the manual definition of such features is time-consuming and heavily reliant on human expertise and domain knowledge. On the other hand, dynamic analysis tools—particularly penetration testing—require a diverse set of test cases crafted by experts. As a result, they also depend heavily on the experience and intuition of security professionals to identify vulnerable areas within a software system \cite{b5}.

Recognizing the limitations of traditional methods, and with the increasing availability of open-source software repositories, researchers have advocated for a data-driven approach to software vulnerability detection. Accordingly, various machine learning (ML) techniques have been applied to learn features indicative of vulnerabilities and to automate the identification process, with varying degrees of success \cite{b6, b7, b8, b9, b10, b11, b12, b13, b14, b15, b16, b17, b18, b24, b25}.

Much of the existing research has focused on detecting vulnerabilities in languages such as C \cite{b19}. However, despite Python’s widespread adoption in modern software development, it has received relatively little attention in the context of vulnerability detection. Only a limited number of studies \cite{b11, b13, b18, b20, b21, b22, b24, b45, b46} specifically target vulnerabilities in Python source code. Among these, only a few \cite{b13, b18} explore the impact of different code representation techniques—also known as code embeddings—on model performance.

Since machine learning models require numerical input, source code must first be transformed into a structured numeric format before it can be effectively analyzed. It is well established that the quality and relevance of data representation significantly influence the performance of predictive models. However, as there are relatively few studies in this area \cite{b18}, substantial opportunities remain for further experimentation particularly through the evaluation of a wider range of embedding techniques and their integration with various deep learning architectures.

In this work, our goal is to identify the most effective combination of code embedding models and machine learning classifiers for vulnerability detection in Python source code. To this end, we evaluate three code embedding models—Word2Vec, CodeBERT, and GraphCodeBERT—in combination with two deep learning classification models, i.e., BiLSTM and CNN. Our objective is to determine which code embedding-classifier combination delivers the best performance for this task.

In this area for the context of Python, \cite{b18} employs Word2Vec, FastText, and BERT with LSTM-based classifiers. Similarly, \cite{b13} uses Word2Vec, FastText, and CodeBERT with six classifiers, including XGBoost, Multilayer Perceptron (MLP), CNN, LSTM, and Gated Recurrent Unit (GRU). In contrast to these studies, our approach incorporates GraphCodeBERT and evaluates it in combination with BiLSTM and CNN to assess its effectiveness in detecting vulnerabilities in Python source code. A detailed comparison with these related works is presented in Table-1. Our results show that BiLSTM with Word2Vec is the superior combination among all evaluated models, achieving an average precision of 96.2\%, average recall of 93.3\%, average F-score of 94.7\%, and average accuracy of 98.6\%.  It is also noteworthy that CNN with GraphCodeBERT demonstrates strong performance; however, the overall best results are achieved by the BiLSTM with Word2Vec combination. The choice of embedding plays a critical role in the performance of deep learning models, and selecting the appropriate combination of embeddings and classifiers can significantly enhance vulnerability detection outcomes.

Section 2 reviews related work, Section 3 presents the experimental design and performance evaluation, and Section 4 concludes the study and discusses future work.
\section{Related Work}

Research on source code vulnerability detection has gained increasing attention due to the escalating threat of cybercrimes. The focus is shifting from traditional static and dynamic analysis methods \cite{b3, b4, b23} towards machine learning and deep learning approaches \cite{b6, b7, b8, b9, b10, b11, b12, b13, b14, b15, b16, b17, b18}. Sharma et al. \cite{b14} provide a comprehensive review of machine learning techniques applied to source code analysis, while Semasaba et al. \cite{b15} conduct a systematic literature review on software vulnerability detection using deep learning models.

However, a limited number of studies \cite{b11, b13, b18, b20, b21, b22} focus specifically on vulnerabilities in Python source code. Among these, only a few \cite{b13, b18} have analyzed the effectiveness of different code embedding methods for enhancing vulnerability detection performance.

 In the context of Python, Alfadel et al. \cite{b20} empirically analyze security vulnerabilities in Python packages. Harzevili et al. \cite{b21} focus on detecting vulnerabilities within Python machine learning libraries. Ehrenberg et al. \cite{b22} apply named entity recognition techniques to identify key elements in Python source code. Wartschinski et al. \cite{b11} utilize Word2Vec embeddings combined with Long Short-Term Memory (LSTM) networks for vulnerability detection in Python code.

Our work is closely related to that of Bagheri and Hegedűs \cite{b18}, who compare code embedding methods—Word2Vec, FastText, and BERT—combined with an LSTM classifier for vulnerability prediction in Python. Their results indicate that BERT embeddings paired with LSTM achieve the highest accuracy. However, their focus is primarily on evaluating embedding techniques rather than developing a state-of-the-art prediction model.

Similarly, Wang et al. \cite{b13} evaluate three embeddings—Word2Vec, FastText, and CodeBERT—alongside six classifiers: Random Forest, XGBoost, Multilayer Perceptron (MLP), Convolutional Neural Network (CNN), Long Short-Term Memory (LSTM), and Gated Recurrent Unit (GRU). Their findings indicate that Word2Vec outperforms the other embeddings in terms of precision, recall, and F-score. Among the classifiers, LSTM and GRU achieve strong results, with Bi-directional LSTM and GRU enhanced by attention mechanisms (using Word2Vec) yielding the best performance for Python vulnerability detection.

Building on these studies \cite{b13, b18}, our research incorporates Word2Vec and CodeBERT embeddings while extending the investigation to include GraphCodeBERT—a model designed to capture both syntactic and semantic structures of source code. A detailed comparison with related work is presented in Table 1. The main aim of this study is to determine which combination of code embedding and classifier delivers the best performance for vulnerability detection in Python source code. Our experiments reveal that, although CodeBERT and GraphCodeBERT are the latest embedding methods and CNN combined with GraphCodeBERT shows strong performance, the BiLSTM model paired with traditional Word2Vec consistently outperforms all other experimented combinations, making it a reliable choice for this task.

{\tabcolsep=5pt\def\arraystretch{1.3}
\begin{table*}
\centering
\renewcommand{\arraystretch}{1.4}

\begin{tabular}{|p{3.1cm}|p{2.7cm}|p{2.2cm}|p{2.7cm}|p{3cm}|}
\hline
\multirow{1}{3cm}{\textbf{Approaches}} & \multirow{1}{1.2cm}{\textbf{Metrics}} & \textbf{Bagheri and Hegedűs\cite{b18}} & \textbf{Wang et al.  \cite{b13}}& \textbf{Our Work} \\
\hline
\makecell{\textbf{Vulnerabilities}} & SQL Injection & \cmark & \cmark  & \cmark  \\ \cline{2-5}
& XSS & \cmark & \cmark  & \cmark  \\ \cline{2-5}
& Command Injection & \cmark & \cmark  & \cmark  \\ \cline{2-5}
& XSRF & \cmark & \cmark  & \cmark \\ \cline{2-5}
& Remote Code Execution & \cmark & \cmark  & \cmark \\ \cline{2-5}
& Path Disclosure & \cmark & \cmark  & \cmark \\ \cline{2-5}
& Open Redirect & \xmark & \cmark  & \cmark \\ \cline{1-5}

\multirow{5}{*}{\hspace{0.5cm}\makecell{\textbf{Code}\\ \textbf{Embeddings}}} & Word2Vec & \cmark & \cmark  & \cmark \\ \cline{2-5}
& fastText & \cmark & \cmark & \xmark  \\ \cline{2-5}
& CodeBert & \cmark  & \cmark  & \cmark \\ \cline{2-5}
& GraphCodeBert & \xmark & \xmark  & \cmark \\ \cline{1-5}

\multirow{5}{*}{\hspace{0.5cm}\makecell{\textbf{Deep}\\\textbf{Learning}\\ \textbf{Models}}} & LSTM & \cmark & \cmark  & \xmark \\ \cline{2-5}
& BiLSTM & \xmark & \cmark & \cmark  \\ \cline{2-5}
& CNN & \xmark  & \xmark  & \cmark \\ \cline{2-5}
& MLP  & \xmark & \cmark  & \cmark in our previous work\cite{b24} \\ \cline{2-5}
& GRU  & \xmark & \cmark  & \xmark \\ \cline{1-5}

\multirow{4}{*}{\hspace{0.4cm}\textbf{Performance}}& Average Precision & (LSTM+BERT) 91.4\% & (LSTM+Word2Vec) 89.5\% &(BiLSTM+Word2Vec) \textbf{96.2\%}\\ \cline{2-5}
& Average Recall & (LSTM+BERT) 83.2 & (LSTM+Word2Vec) 80.5\% &(BiLSTM+Word2Vec) \textbf{93.3\%}\\ \cline{2-5}
& Average F-Score & (LSTM+BERT) 87.1\% &(LSTM+Word2Vec) 84.6\% & (BiLSTM+Word2Vec) \textbf{94.7\%}\\ \cline{2-5}
& Average Accuracy & (LSTM+BERT) 93.8\% &- &(BiLSTM+Word2Vec) \textbf{98.6\%}\\ \cline{1-5}

\end{tabular}
\caption{Comparison of our work with Related Work}
\end{table*}}

\section{Experimental Design}
This section demonstrates the details related to the experimental study incorporating types of selected vulnerabilities, dataset and its preprocessing, code embeddings, machine learning algorithms, and their performance evaluations. All experiments are executed on Windows 11 on Dell OptiPlex 5000 SFF with processor Intel core i5-12500, RAM 32 GB (2*16GB) DDR4 having AMD® Radeon™ 550, 2 GB GDDR5 graphic card.

\subsection{Selection of Vulnerabilities}
We examine the same software vulnerabilities highlighted in Bagheri and Hegedűs \cite{b18} and Wang et al.\ \cite{b13}, namely SQL Injection, Cross-Site Scripting (XSS), Command Injection, Cross-Site Request Forgery (XSRF), Path Disclosure, Remote Code Execution, and Open Redirect. A brief description of each vulnerability type is provided below:

SQL Injection: An attacker inserts malicious code into SQL commands through web pages, aiming to gain unauthorized access to the database or execute commands illicitly.

Cross-Site Scripting (XSS): An attacker injects malicious scripts into legitimate web pages or applications, causing the victim’s browser to execute these scripts, potentially compromising user data or session information.

Command Injection: An attack where the adversary executes arbitrary commands on the host operating system via a vulnerable application.

Cross-Site Request Forgery (XSRF): An attack that tricks authenticated users into submitting unwanted requests to a web application in which they are currently authenticated.

Remote Code Execution: An attacker exploits a vulnerability to remotely execute arbitrary code on a target system over a network.

Open Redirect: An attacker manipulates URLs to redirect users from a legitimate website to a malicious site.

\subsection{Dataset and Preprocessing}
We use the dataset prepared by Wartschinski et al.\ \cite{b11}, available at \cite{b42, b44}, which is compiled by targeting publicly accessible GitHub repositories. GitHub stands out as the largest repository hosting platform for source code globally, making it an ideal resource for this work.

GitHub has a version control system that revolves around commits, which allows users to keep track of their work and explore the changes/updates made on coding scripts over time. A commit that fixes a bug or vulnerability can be described
as a patch, consisting of a pair of software versions, a buggy or a flawed one, and an updated and (hopefully) the correct one. By analyzing the differences between the old and the new versions, vulnerable code patterns can be learned. 

Wartschinski et al.\ \cite{b11} gather a distinct dataset for each vulnerability type. The data is collected in the form of commits that contain security-related fixes. Sections of code that are updated or removed in these commits are categorized as vulnerable, along with the surrounding code to provide context. Conversely, the remaining code and the post-fix version are labeled (probably) as not vulnerable. We use 70\% data in training, 15\% in testing, and 15\% in the validation set.

\begin{table*}
	\centering
	\renewcommand{\arraystretch}{1.6}
	\begin{tabular}{|p{4.0cm}|c|c|c|c|c|c|}
		\hline
		
		\textbf{Github Repositories}& \textbf{URL}  \\ 
		\hline
		Numpy&https://github.com/numpy/numpy \\ \cline{1-2}
		Django&https://github.com/django/django\\ \cline{1-2}
		Sci-kit-learn&https://github.com/scikit-learn/scikit-learn\\ \cline{1-2}
		Tensorflow&https://github.com/tensorflow/tensorflow\\ \cline{1-2}
		Scipy&https://github.com/scipy/scipy\\ \cline{1-2}
		Flask&https://github.com/pallets/flask\\ \cline{1-2}
		Docker  Compose&https://github.com/docker/compose\\ \cline{1-2}
		eBPFCat&https://github.com/tecki/ebpfcat\\ \cline{1-2}
		
	\end{tabular}
	\caption{Github repositories used in the training of Word2Vec model}
\end{table*} 
\subsection{Code Embeddings}
\begin{table*} 
\centering
		\renewcommand{\arraystretch}{1.5}

		\begin{tabular}{|p{2.8cm}|p{2cm}|p{2.2cm}|p{2.2cm}|p{2.3cm}|c|c|c|c|}
			\hline
			\multirow{1}{3cm}{\textbf{Vulnerabilities}} & \multirow{1}{1.2cm}{\textbf{Metrics}} & \multicolumn{1}{c|}{\textbf{Word2Vec}}& \multicolumn{1}{c|}{\textbf{CodeBERT}} & \multicolumn{1}{c|}{\textbf{GraphCodeBERT}} \\ \cline{2-5}
			
			\hline{}
			\textbf{SQL injection}& Precision & \textbf{96.8\%} & 66.2\%  & 72.8\%   \\ \cline{2-5}
			& Recall & \textbf{93.8\%} &46.7\%  & 79.5\%    \\ \cline{2-5}
			& F-Score & \textbf{95.3\%} &54.8\%  & 76.0\%    \\ \cline{2-5}
			& Accuracy & \textbf{98.2\%} &85.6\%  & 90.6\%    \\ \cline{1-5}
			
			\textbf{XSS}& Precision  & \textbf{94.8\%} & 94.6\%  & 89.2\%   \\ \cline{2-5}
			& Recall  & 91.3\% & 68.0\%  & \textbf{95.7\%}   \\ \cline{2-5}
			& F-Score  & \textbf{93.0\%} & 79.1\%  & 92.3\%   \\ \cline{2-5}
			& Accuracy  & \textbf{98.8\%} & 96.8\%  & 98.6\%   \\ \cline{1-5}

			\textbf{Command injection}& Precision  & \textbf{97.8\%} & 92.3\%  & 83.6\%   \\ \cline{2-5}
			& Recall & \textbf{95.7\%} & 57.2\%  & 88.0\%   \\ \cline{2-5}
			& F-Score   & \textbf{96.7\%} & 70.6\%  & 85.8\%   \\ \cline{2-5}
			& Accuracy  & \textbf{99.1\%} & 93.8\%  & 96.3\%   \\ \cline{1-5}

			\textbf{XSRF}& Precision  & \textbf{96.7\%} & 72.6\%  & 84.6\%   \\ \cline{2-5}
			& Recall  & \textbf{90.7\%} & 52.5\%  & 82.7\%   \\ \cline{2-5}
			& F-Score  & \textbf{93.6\%} & 60.9\%  & 83.6\%   \\ \cline{2-5}
			& Accuracy  & \textbf{98.3\%} & 91.2\%  & 95.5\%   \\ \cline{1-5}

			\textbf{Remote code execution}& Precision & \textbf{97.2}\% & 68.4\%  & 94.5\%   \\ \cline{2-5}
			& Recall  & \textbf{95.9\%} & 81.0\%  & 88.2\%   \\ \cline{2-5}
			& F-Score  & \textbf{96.5\%} & 74.2\%  & 91.3\%   \\ \cline{2-5}
			& Accuracy  & \textbf{99.4\%} & 95.3\%  & 98.5\%   \\ \cline{1-5}

			\textbf{Path disclosure}& Precision  & \textbf{97.7\%} & 81.2\%  & 79.0\%   \\ \cline{2-5}
			& Recall  & \textbf{96.9\%} & 66.1\%  & 84.5\%   \\ \cline{2-5}
			& F-Score  & \textbf{97.3\%} & 72.8\%  & 81.7\%   \\ \cline{2-5}
			& Accuracy  & \textbf{99.3\%} & 94.7\%  & 95.6\%   \\ \cline{1-5}

			\textbf{Open redirect}& Precision& \textbf{92.5}\% & 86.6\%  & 77.0\%   \\ \cline{2-5}
			& Recall  & \textbf{89.0\%} & 55.2\%  & 86.5\%   \\ \cline{2-5}
			& F-Score  & \textbf{90.7\%} & 67.5\%  & 81.5\%   \\ \cline{2-5}
			& Accuracy  & \textbf{97.5\%} & 93.0\%  & 94.2\%   \\ \cline{1-5}

		\end{tabular}
		\caption{BiLSTM Results with Code Embeddings}
		
\end{table*}
\begin{table*} 
\centering
		\renewcommand{\arraystretch}{1.5}

		\begin{tabular}{|p{2.8cm}|p{2cm}|p{2.2cm}|p{2.2cm}|p{2.3cm}|c|c|c|c|c|}
			\hline
			\multirow{1}{3cm}{\textbf{Vulnerabilities}} & \multirow{1}{1.2cm}{\textbf{Metrics}} & \multicolumn{1}{c|}{\textbf{Word2Vec}}& \multicolumn{1}{c|}{\textbf{CodeBERT}} & \multicolumn{1}{c|}{\textbf{GraphCodeBERT}} 
			\\ \cline{2-5}
			
			\hline{}
			\textbf{SQL injection}& Precision & \textbf{93.5}\% & 81.1\%  & 93.2\%   \\ \cline{2-5}
			& Recall & 83.2\% &76.9\%  & \textbf{91.8\%}    \\ \cline{2-5}
			& F-Score & 88.1\% &79.0\%  & \textbf{92.5\%}    \\ \cline{2-5}
			& Accuracy & \textbf{95.9\%} &87.4\%  & 95.4\%    \\ \cline{1-5}
			
			\textbf{XSS}& Precision  & 97.4\% & 96.2\%  & \textbf{98.0\%}   \\ \cline{2-5}
			& Recall  & 75.9\% & 87.2\%  & \textbf{91.9\%}   \\ \cline{2-5}
			& F-Score  & 85.3\% & 91.5\%  & \textbf{94.8\%}   \\ \cline{2-5}
			& Accuracy  & 97.6\% & 97.4\%  & \textbf{98.3\%}   \\ \cline{1-5}

			\textbf{Command injection}& Precision  & 96.3\% & \textbf{96.8\%}  & 95.6\%   \\ \cline{2-5}
			& Recall & 86.6\% & \textbf{91.9\%}  & 87.5\%   \\ \cline{2-5}
			& F-Score   & 91.2\% & \textbf{94.3\%}  & 91.4\%   \\ \cline{2-5}
			& Accuracy  & 95.0\% & 97.5\%  &  \textbf{98.0\%}   \\ \cline{1-5}

			\textbf{XSRF}& Precision  & 91.3\% & 83.6\%  &\textbf{94.3\%}   \\ \cline{2-5}
			& Recall  & 83.3\% & 81.8\%  & \textbf{93.1\%}   \\ \cline{2-5}
			& F-Score  & 87.1\% & 82.7\%  & \textbf{93.7\%}   \\ \cline{2-5}
			& Accuracy  & 97.2\% & 92.7\%  & \textbf{97.3\%}   \\ \cline{1-5}

			\textbf{Remote code execution}& Precision & \textbf{97.4}\% & 93.1\%  & 94.4\%   \\ \cline{2-5}
			& Recall  & 89.5\% & 88.9\%  & \textbf{94.9\%}   \\ \cline{2-5}
			& F-Score  & 93.2\% & 91.0\%  & \textbf{94.6\%}   \\ \cline{2-5}
			& Accuracy  & \textbf{98.9\%} & 97.1\%  & 98.2\%   \\ \cline{1-5}

			\textbf{Path disclosure}& Precision  & \textbf{94.9\%} & 94.6\%  & 93.1\%   \\ \cline{2-5}
			& Recall  & 85.3\% & \textbf{91.0\%}  & 90.4\%   \\ \cline{2-5}
			& F-Score  & 89.9\% & \textbf{92.7\%}  & 91.7\%   \\ \cline{2-5}
			& Accuracy  & 95.0\% & 96.9\%  & \textbf{96.0\%}   \\ \cline{1-5}

			\textbf{Open redirect}& Precision& 92.0\% & \textbf{95.8}\%  & 92.8\%   \\ \cline{2-5}
			& Recall  & 91.5\% & 90.7\%  &  \textbf{94.0\%}   \\ \cline{2-5}
			& F-Score  & 91.8\% & 93.2\%  & \textbf{94.4\%}   \\ \cline{2-5}
			& Accuracy  & 97.9\% & 97.1\%  & \textbf{98.0\%}   \\ \cline{1-5}

		\end{tabular}
		\caption{CNN Results with Code Embeddings}
\end{table*}

For training machine learning algorithms, it is essential to represent code tokens as vectors that preserve relevant semantic and syntactic information. In this study, we employ three code embedding techniques: Word2Vec, CodeBERT, and GraphCodeBERT.

\textbf{Word2Vec Embeddings:}

We use the Python GitHub repositories listed in Table-2 to construct a high-quality corpus for training the Word2Vec model. This model is employed to encode Python code tokens into continuous vector representations. We adopt the same hyperparameter ranges suggested in \cite{b11}, where the number of training iterations varies between 1 and over 100, vector dimensions range from 5 to 300, and the minimum token count ranges from 10 to 500. After experimentation, we obtain the best results using 200 training iterations, a minimum count of 10, and a vector dimension of 300. These embeddings serve as feature vectors for downstream deep learning classifiers tasked with vulnerability detection.

\textbf{CodeBert Embeddings:}

CodeBERT is a pre-trained model based on the Bidirectional Encoder Representations from Transformers (BERT) architecture, specifically designed for both programming and natural language tasks. Trained on large-scale datasets that include paired source code and natural language, it captures both syntactic and semantic features of code. In this study, we utilize the pretrained microsoft/codebert-base model from the HuggingFace Transformers library to extract contextualized embeddings for Python source code. Each code snippet is tokenized, truncated if necessary to fit within the model’s input limits, and augmented with special tokens such as [CLS], [SEP], and [EOS] to match the model’s expected input format. The embeddings are obtained from the model’s final hidden layer (last\_hidden\_state) and represent each token in the code sequence. These token-level embeddings are then used as feature vectors for downstream deep learning classifiers for the task of vulnerability detection.

\textbf{GraphCodeBert Embeddings:}

We utilize the pretrained GraphCodeBERT model, an extension of CodeBERT that integrates code structure information such as data flow and syntax graphs to capture richer semantic relationships within source code. This model is specifically designed to better represent programming language nuances compared to traditional token-based models. Using the microsoft/graphcodebert-base pretrained weights, we tokenize Python source code, truncate it to a maximum length, and add special tokens [CLS], [SEP], and [EOS]. The model outputs contextualized token embeddings from the last hidden layer, which serve as feature vectors for downstream deep learning classifiers tasked with vulnerability detection. 

\subsection{Machine Learning Algorithms}
In this study, we employ two deep learning algorithms, i.e., Bidirectional Long Short-Term Memory (BiLSTM) and Convolutional Neural Network (CNN). Both models utilize the generated code embeddings to learn meaningful representations essential for accurately identifying vulnerabilities in Python source code.

\textbf{Bidirectional Long Short-Term Memory} (BiLSTM) in contrast with LSTM, processes sequential data in both forward and backward directions using two separate hidden layers, enabling enhanced training by traversing the input sequence twice. We implement our BiLSTM model using the Python Keras framework with TensorFlow as the backend.

\textbf{Selection of optimal hyper-parameters for BiLSTM:} BiLSTM networks offer extensive configurability through various hyperparameters, and selecting the right combination is vital for optimal model performance. We experimented with tuning several parameters, including the number of BiLSTM layers, optimizer, learning rate, loss function, number of training epochs, and batch size. Our experiments revealed that the following setup delivers strong results: one input layer, three hidden layers each consisting of 50 neurons (with BiLSTM processing data in both directions by duplicating and concatenating hidden states), four dropout layers set at a rate of 0.2 to prevent overfitting, and a single-node output layer. We use Adam optimizer and mean\_squared\_error as the loss function. The model is trained for 50 epochs with a batch size of 128, employing early stopping based on validation loss to prevent overfitting.

\textbf{Convolutional Neural Network (CNN)} is a deep learning architecture well-suited for extracting spatial features from structured data. CNNs utilize convolutional layers with learnable filters that scan input data to capture local patterns and hierarchical representations. We build our CNN model using the Python Keras framework with TensorFlow as the backend.

\textbf{Selection of optimal hyperparameters for CNN:}
Similar to BiLSTM, CNN also offers extensive configurability through several hyperparameters. Our CNN architecture begins with a downscaling convolutional block composed of a 1D convolutional layer with 32 filters and kernel size 5, applied with stride 2 and ‘same’ padding, followed by batch normalization, ReLU activation, and max-pooling to reduce the sequence length. This is followed by three feature extraction blocks with convolutional layers having 64, 128, and 256 filters respectively, each with kernel size 3, batch normalization, ReLU activations, and max-pooling layers that progressively downscale the feature maps. After the convolutional layers, a global average pooling layer summarizes the extracted features across the sequence. The classification head consists of a fully connected dense layer with 200 neurons activated by ReLU, followed by a dropout layer with a rate of 0.2 to prevent overfitting, and a final sigmoid neuron for binary classification output.

Training is performed using the Adam optimizer over 20 epochs with a batch size of 128. To address class imbalance, balanced class weights are computed and applied during training. We utilize callbacks including early stopping based on F1 score to avoid overfitting, and learning rate reduction on plateau to improve convergence. 
\begin{figure*}
	\centering
         \subfigure[XSS (Model BiLSTM)]{
		\includegraphics[width=58mm, height=6cm]{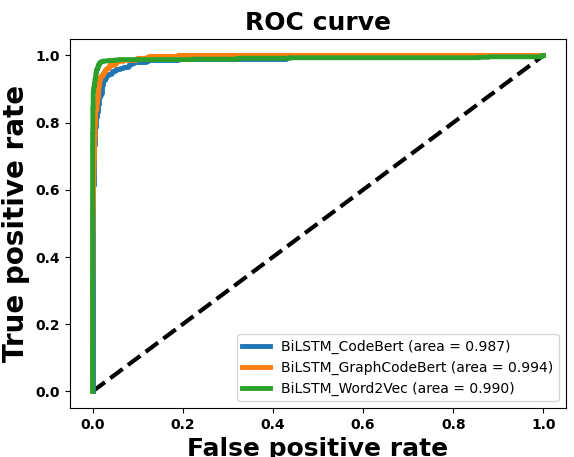}
	}
	\subfigure[SQL Injection (Model CNN)]{
		\includegraphics[width=58mm, height=6cm]{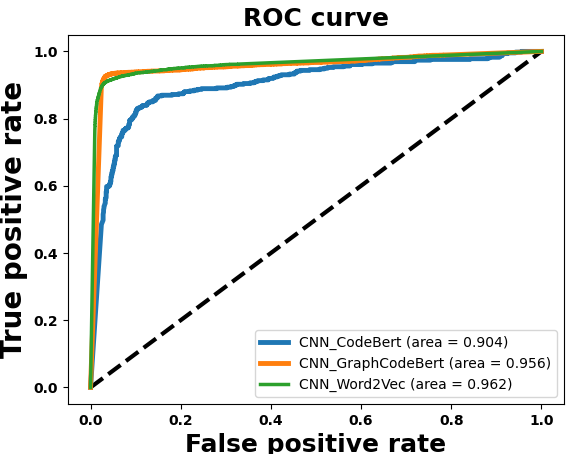}
	}
	\subfigure[XSRF (Model CNN)]{
		\includegraphics[width=58mm, height=6cm]{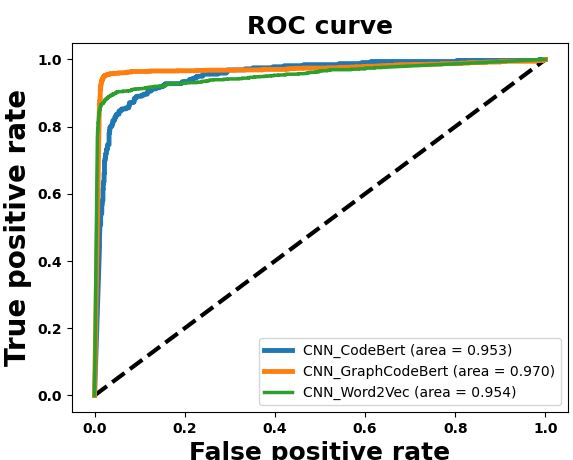}
	}

	\caption{Receiver Operating Characteristic (ROC) of Models}
	
\end{figure*}

\subsection{Performance Evaluation} Among the three embedding techniques evaluated—Word2Vec, CodeBERT, and GraphCodeBERT—the BiLSTM model with Word2Vec embeddings demonstrates superior performance across all key metrics. It achieves an average precision of 96.2\%, average recall of 93.3\%, average F-score of 94.7\%, and average accuracy of 98.6\%. In comparison, the CNN model integrated with GraphCodeBERT shows the best performance among CNN-based models across the same embeddings, attaining an average precision of 94.4\%, average recall of 91.9\%, average F-score of 93.3\%, and average accuracy of 97.3\%. However, when directly comparing BiLSTM with Word2Vec and CNN with GraphCodeBERT, the former exhibits moderately better performance. While the difference is not highly significant, even slight improvements can be meaningful. It is also worth noting that the other combinations did not perform poorly, but the subtle differences emphasize the importance of choosing the most effective model. Detailed results are presented in Tables 4 and 5, while Figure 1 illustrates the ROC curves for some selected vulnerability types.

\section{Conclusion and Future Work} 
In this study, we investigate the impact of different code embedding models, i.e., Word2Vec, CodeBERT, and GraphCodeBERT combined with deep learning classifiers BiLSTM and CNN for the task of vulnerability detection in Python source code. Our experimental results demonstrate that the combination of BiLSTM with Word2Vec consistently outperforms all other configurations in terms of precision, recall, F-score, and accuracy. Notably, CNN with GraphCodeBERT also shows strong performance, ranking second overall. While the remaining combinations do not perform very poorly, even relatively small performance differences can be significant when aiming for optimal results. Hence, selecting the most effective code embedding-classifier combination is crucial for advancing automated vulnerability detection in software. For future work, we plan to explore additional machine learning models combined with code embeddings and evaluate their effectiveness specifically for vulnerability detection in Python source code.



\end{document}